\documentclass[preprint,authoryear,12pt]{elsarticle}
\usepackage{amssymb}
\usepackage{amsmath}
\usepackage{graphicx}
\usepackage{tabularx}
\usepackage{tabulary}
\usepackage{multirow}
\usepackage{appendix}
\usepackage{verbatim}
\usepackage[nodots]{numcompress}
\usepackage{xcolor}
\usepackage{listings}
\usepackage{algorithm}
\usepackage{algorithmic}
\usepackage{url}
\newcolumntype{Y}{>{\centering\arraybackslash}X}
\journal{Digital Investigation}

\begin{document}

%%% [ARXIV]

\begin{center}
{\Huge Forensic Analysis of the ChatSecure Instant Messaging Application on Android Smartphones}
\end{center}

\begin{center}
{\Large%\centering
\ \\
Please, cite this paper as:\\{\bfseries
%\flushleft
%\begin{tabular}{c}
Cosimo Anglano, Massimo Canonico, Marco Guazzone,\\
\emph{``Forensic Analysis of the ChatSecure Instant Messaging Application on Android Smartphones,''}\\
Digital Investigation, Volume 19, December 2016, Pages 44--59.\\
DOI:\url{10.1016/j.diin.2016.10.001}\\
%{\normalsize Publisher: \url{http://www.sciencedirect.com/science/article/pii/S1742287616300950}}
{\normalsize Publisher: \url{http://dx.doi.org/10.1016/j.diin.2016.10.001}}
%\end{tabular}
}}
\end{center}

\newpage

\begin{frontmatter}	
%\title{Forensic Analysis of the ChatSecure Instant Messaging Application on Android Smartphones}
\title{Forensic Analysis of the ChatSecure Instant Messaging Application on Android Smartphones\tnoteref{cite,pub}}

\tnotetext[cite]{Please, cite as: \emph{Cosimo Anglano, Massimo Canonico, Marco Guazzone, ``Forensic Analysis of the ChatSecure Instant Messaging Application on Android Smartphones,'' Digital Investigation, Volume 19, December 2016, Pages 44--59, DOI: 10.1016/j.diin.2016.10.001}}
%\tnotetext[pub]{Link to publisher: \url{http://www.sciencedirect.com/science/article/pii/S1742287616300950}}
\tnotetext[pub]{Link to publisher: \url{http://dx.doi.org/10.1016/j.diin.2016.10.001}}

\author[add1]{Cosimo Anglano\corref{corauth}}
\ead{cosimo.anglano@uniupo.it}
\author[add1]{Massimo Canonico}
\ead{massimo.canonico@uniupo.it}
\author[add1]{Marco Guazzone}
\ead{marco.guazzone@uniupo.it}

\address[add1]{DiSIT - Computer Science Institute,\\ University of Piemonte Orientale, Alessandria (Italy)}
\cortext[corauth]{Corresponding author. Address: viale T. Michel 11, 15121 Alessandria (Italy). Phone: +39 0131 360188.}

\address{(Please, cite as:\\ \textbf{Cosimo Anglano, Massimo Canonico, Marco Guazzone,\\
\emph{``Forensic Analysis of the ChatSecure Instant Messaging Application on Android Smartphones,''}\\
Digital Investigation, Volume 19, December 2016, Pages 44--59. DOI: 10.1016/j.diin.2016.10.001})}

%%% [/ARXIV]

\begin{abstract}
We present the forensic analysis of the artifacts
generated on Android smartphones by \textit{ChatSecure}, a \emph{secure} 
Instant Messaging application that provides strong encryption for transmitted and 
locally-stored data to ensure the privacy of its users.

We show that ChatSecure stores local copies of both exchanged
messages and files into two distinct, AES-256 encrypted databases,
and we devise a technique able to decrypt them when the secret passphrase, chosen
by the user as the initial step of the encryption process, is known.

Furthermore, we show how this passphrase can be identified and extracted
from the volatile memory of the device, where it persists  for the entire execution of ChatSecure
after having been entered by the user,
thus allowing one to carry out decryption even if the passphrase is
not revealed by the user.

Finally, we discuss how to analyze and correlate the
data stored in the databases used by ChatSecure to identify the
IM accounts used by the user and his/her buddies to communicate,
as well as to reconstruct the chronology and contents 
of the messages and files that have been  exchanged among them.

For our study we devise and use an experimental methodology, based on the use of emulated devices,
that provides a very high degree of reproducibility of the results, and we validate the
results it yields against those obtained from real smartphones.
\end{abstract}


\end{frontmatter}

\section{Introduction}
\label{intro}
\emph{Instant Messaging} (\emph{IM}) applications are very popular among smartphone users 
because of the level of convenience they provide in interpersonal communications.
Quite sophisticated IM applications are available today for the prominent smartphone 
platforms (e.g., Android, iOS, and Windows Phone, to name a few) that
allow users to exchange text and files in (pseudo) real time.

In addition to legitimate uses, however, IM applications 
are increasingly being used to carry out illicit 
activities~\citep{unodc-2013}.
Therefore, the forensic analysis of 
these applications has received considerable attention in the recent past.
IM-based evidence may indeed prove crucial in all those cases where an IM 
application has been used by the parties involved in a crime, i.e.\ by a 
perpetrator to interact with its victims, or by criminals in the attempt to 
escape interception when they communicate.

Generally speaking, the forensic analysis of an IM application is based on the 
availability  of various types of artifacts (metadata and content of exchanged messages and files, log files, etc.)
stored by that application on the local storage of smartphones. By 
locating, extracting, and analyzing such artifacts, quite often it is possible 
to recover a significant amount of potential 
evidence~\citep{social-im-forensics,anglano-whatsapp,iforensics}.

This situation, however, is rapidly changing.
The increasing awareness of the fact that IM communications may be 
intercepted when transiting over the infrastructure of the service provider, is 
stimulating the interest towards \emph{secure} IM 
applications~\citep{chatsecure,telegram,textsecure,gliph,wickr}. These 
application, unlike standard ones, provide suitable 
privacy-preserving and user security 
mechanisms, such as strong encryption for transmitted and locally-stored data, 
secure user authentication, plausible deniability, forward secrecy, and so on.
Secure IM applications pose new challenges to the forensic analyst, that has to 
deal with the issues posed by the privacy-preserving features of these 
applications.

Among these applications, one that is receiving increasing attention
is \emph{ChatSecure}~\citep{chatsecure}, which is available both for Android and 
iOS.
There are various reasons for its success: (a) it is open-source (so it is 
possible to audit its code), (b) it provides message encryption, partner 
authentication, deniability and perfect forward secrecy thanks to the use of 
the \emph{Off-The-Record (OTR)}~\citep{otr} messaging system (which has gained 
an excellent reputation in the privacy-concerned user communities), (c) it 
encrypts 
locally stored data with \emph{SQLCipher}~\citep{sqlcipher} and
\emph{IOCipher}~\citep{iocipher}, and (d) it has been 
ranked as one of the most secure IM applications by the 
\emph{Electronic Frontiers Foundation (EFF)}~\citep{secureim-eff}.

Given these characteristics, the interest on the forensic analysis of 
ChatSecure is evident, since there 
is no publicly known way of decrypting OTR-encrypted data 
once they are in transit over the network. Thus, forensic analysis of the 
devices used to communicate with ChatSecure may be the only option available to 
retrieve IM-based evidence.

To the best of our knowledge, there is no published work addressing the 
forensic analysis of ChatSecure on the Android platform.
In this paper we fill this gap by describing which artifacts ChatSecure
are stored in the local memory of the device, and how they 
can be decoded and correlated among them to infer information of potential 
investigative interest.

The original contributions of this paper can be summarized as follows:
\begin{itemize}
	\item we show that ChatSecure stores locally copies of all the 
	messages and files that are exchanged between the user and her contacts
	into two encrypted \emph{SQLite v.3}~\citep{sqlite3} databases;
	\item we analyze the encryption procedure used for these databases, and
	we develop and implement an algorithm able to decrypt them using the secret 
	passphrase set by the user;
	\item we show how the passphrase can be retrieved
	from the volatile memory of an Android device;
	\item we discuss the decoding and the interpretation of all the artifacts generated
	by ChatSecure, and we show how they can be correlated to perform
	various forensic reconstructions, such as the chronology and contents of exchanged files
	and messages, the set of IM accounts used by the ChatSecure user, as well as
	the list of contacts associated with each one of them;
	\item we show that it is not possible to recover the information deleted by ChatSecure users 
	because of the use of secure deletion techniques in SQLCipher and IOCipher;
	\item we devise and use an experimental methodology, based on the use of emulated devices,
	that provides a very high degree of reproducibility of the results, and we validate the
	results it yields against those obtained from real smartphones.
\end{itemize}

The rest of the paper is organized as follows. In Sec.~\ref{related}
we review existing work, while in Sec.~\ref{methodology} we describe
the methodology and the tools we use in our study.
Then, in Sec.~\ref{chatsecure} we discuss the forensic analysis of
ChatSecure and, in Sec.~\ref{conclusions}, we
conclude the paper. 
\section{Related works}
\label{related}
Smartphone forensics has been widely studied in the recent literature, 
which mostly focuses on Android and iOS 
forensics~\citep{learning-android-forensics,learning-ios-forensics},
given the pervasiveness of these platforms.
As a result, well known and widely accepted methodologies and 
techniques are available today that are able to properly deal with the 
extraction and analysis of evidence from smartphones.
In this paper we leverage this vast body of work for extracting and analyzing 
the data generated by ChatSecure during its usage.

The importance of the forensic analysis of smartphone IM applications has been 
also acknowledged in the literature, where a significant number of papers on 
this 
topic has been published.
\citep{anglano-whatsapp} discusses the forensic analysis of WhatsApp 
Messenger.
\citep{iforensics} focuses on the forensic analysis of
three IM applications (namely AIM, Yahoo! Messenger, and Google
Talk) on the iOS platform.
\citep{social-im-forensics} presents the analysis of 
several IM applications on various smartphone
platforms, aimed at identifying the encryption
algorithms used by them.
\citep{tso-iphone} discusses the analysis of iTunes backups for
iOS devices aimed at identifying the artifacts left by
various social network applications.
\citep{forensic-android-dfrws2015} discusses the analysis of the data 
transmitted or stored locally by 20 popular Android IM applications.

None of these papers, however, covers the forensic analysis of 
ChatSecure on Android platforms, which is instead the focus of this 
paper. The closest work to ours is \citep{chatsecure:quarkslab}, that
is focused on the iOS platform. However, the results discussed there
do not apply to Android, given the significant differences existing between
the Android version and the iOS version of ChatSecure.
\section{Analysis methodology and tools}
\label{methodology}

The study described in this paper has been performed by
carrying out a set of controlled experiments, each one referring to a specific
usage scenario (one-to-one communication, group communication, file 
exchange, etc.), during which typical user interactions have taken place.
After each experiment, the internal memory of the sending and receiving devices
has been examined in order to identify, decode, and analyze the
data generated by ChatSecure in that experiment.
In all the experiments, we run ChatSecure v.\ 14.2.3 (the last one available on Google Play at
the moment of this writing).

The data generated by ChatSecure are stored into an area of the internal device memory that is normally inaccessible
to users (see Sec.~\ref{artifacts}). Therefore, suitable methodologies and tools need to be
adopted in order to access and acquire this area.
Tools like UFED~\citep{ufed}, XRY~\citep{xry},
and Oxygen Forensics Detective~\citep{oxygen-detective}, among others, are able to
perform this acquisition in a forensically-sound manner.

However, this approach presents some limitations, namely:
\begin{itemize}
	\item \textbf{limited generality}: 
	to gain confidence into the generality of the results,
	a suitably large number of devices and Android versions
	should be used for the experiments.
	The resulting high costs both in terms of purchase and of the
	time required to replicate the experiments on a large set of devices, however, practically
	limits the number of devices used for the experiments, thus potentially casting doubts on
	the generality of the results;
	\item \textbf{limited replicability}: a third party wanting to reproduce the
	results needs to use the same set of devices, operating systems versions,
	and forensic acquisition tools to repeat experiments This, however, may be problematic, both because of
	device availability and of the cost of the acquisition tools;
	\item \textbf{limited controllability}: smartphones are complex devices, running a multitude
	of applications and services, whose behavior and interactions are hard to characterize.
	As a consequence, it may be difficult not only to reproduce the exact conditions
	holding at the moment of each experiment, but also to exclude with certainty
	possible data cross-contaminations among different applications that use the same
	file system (as in Android).
\end{itemize}

To overcome the above limitations, in this work we carry out experiments using
emulated mobile devices instead of physical ones.
In particular, we use the \emph{Android Mobile Device Emulator}~\citep{android-emulator} 
to create various \emph{Android Virtual Devices} (AVDs), that are  emulated smartphones
behaving exactly like real physical devices that can be customized with
different hardware characteristics and Android versions. 
The status of AVDs can be monitored by means of the \emph{Android Device Monitor}~\citep{android-monitor} (ADM).

The use of emulated devices provides many advantages, and allows us to overcome
the limitations discussed above.
First, generality of results is benefited since it is simple and cost-effective
to run experiments on a variety of different AVDs (featuring different hardware and software
combinations), and to quickly extract the contents of their internal memory.
Second, also replicability is greatly benefited, since a third-party can
configure AVDs exactly as we did, thus reproducing the same conditions of our experiments.
Finally, also controllability is enhanced: the configuration of AVDs 
(that include both hardware/software features, as well as a set of services and apps running
in the background) is under total control of the experimenter by means of the ADM,
thus allowing us (as well as a third-party replicating our experiments)
to precisely determine the operational conditions holding on each AVD at the moment
of the experiment.

For our experiments, we use the three AVDs configurations shown in Table~\ref{tab:AVD} below,
that are characterized by different Android versions, 
processor families, and volatile and persistent storage sizes.
\begin{table*}[hbtp]
	\caption{Characteristics of the AVDs used in the experiments.}
	\label{tab:AVD}
	\begin{center}
		\begin{footnotesize}
			\begin{tabularx}{\linewidth}{|c|c|c|X|}
				\hline
				\multicolumn{4}{|c|}{\textbf{Characteristics of AVDs used for experiments}} \\ \hline
				\emph{Processor} & \emph{RAM} (MB) &\emph{Internal storage (MB)} & \emph{Android version}\\ \hline
				ARM (armeabi-v7a) & 512 & 2047 & 4.4 (API 19) \\ \hline
				Intel Atom (x86) & 1536 & 1024 & 5.1 (API 22) \\ \hline
				Intel Atom (x86\_64) & 1536 & 1024 & 6.0 (API 23) \\ \hline
			\end{tabularx}
		\end{footnotesize}
	\end{center}
\end{table*}
To carry out our analysis, we run all the experiments on these AVDs and, at the
end of each experiment, we extract the data generated by ChatSecure using the \emph{pull}
functionality of the \emph{File Explorer}
of the ADM, that allows one to recursively extract
entire folders, or individual files. Alternatively, this task can be carried out using 
the \emph{Android Debug Bridge}~\citep{android-adb} (ADB) to pull data out from the AVD using
a command-line interface.

In order to validate the results obtained with AVDs, we compare them against
results obtained running experiments on a real device.
More precisely, we run experiments on a Samsung SM-G350 Galaxy Core 
Plus smartphone running Android 4.4.2, and we use
the \emph{Cellebrite UFED4PC} platform~\citep{ufed4pc}
to perform device memory extraction, and the UFED \emph{Physical Analyzer}~\citep{ufed_physical}
to decode its contents.
In all the experiments we performed, the results collected from this smartphone
were identical to those obtained from the emulated devices we considered.

In addition to the analysis of the persistent device memory,
we also examine its volatile memory to verify whether encryption keys or secret 
passwords are stored there, and to identify and extract them.
In particular, as discussed in Sec.~\ref{encryption}, we use \textsf{LiME}~\citep{lime} to dump the
contents of the volatile memory of the AVDs used in the experiments,
and \textsf{Volatility}~\citep{volatility} to analyze these dumps.
We perform memory analysis experiments only for the ARM architecture
(row 1 of Table~\ref{tab:AVD}) since, at the moment of this writing, \textsf{LiME} 
supports this architecture only.
Note that in order to work, \textsf{LiME} requires the device to be rooted.
For AVDs, however, this is not an issue, since they are pre-configured to allow
root access to the user.

We do not validate the memory analysis results against real devices,
since \textsf{LiME} requires an Android kernel supporting dynamic module loading.
To enable this functionality, the kernel source must be reconfigured and
recompiled. 
For a real smartphone, the stock Android kernel is not sufficient, as vendors typically customize 
it to suit the specific hardware configuration of the device. Unfortunately, the kernel source for 
the Samsung SM-G350 Galaxy Core Plus smartphone we used for validation was not available to us, 
so we could not configure such a device to work with \textsf{LiME}.

Finally, the source code of ChatSecure (which is freely available from 
\citep{chatsecure_source}) has been examined to verify our hypothesis about its behavior,
or to understand how to decode the data it generates.

For the sake of reproducibility of the experiments we discuss
in this paper, in \citep{lime-howto} we describe how to concretely
configure and use the various tools that we rely upon
to create and run an AVD, and to 
carry out the analysis of its persistent and volatile memory.
\section{Forensic analysis of ChatSecure}
\label{chatsecure}
ChatSecure is an IM application that allows its users to communicate securely via 
their existing accounts on IM providers that use
the XMPP~\citep{xmpp} protocol (e.g.\ \emph{Google Talk} or \emph{Jabber}).

To ensure privacy, ChatSecure provides end-to-end message encryption with OTR and encrypts 
with \emph{SQLCipher}~\footnote{SQLCipher is open-source extension to SQLite that provides 
transparent 256-bit AES encryption of SQLite database files} and 
\emph{IOCipher}~\footnote{IOCipher is a library that implements
encrypted virtual disks using an SQLCipher database (more details can be found in Sec.~\ref{encryption})}
the SQLite databases it uses to store the information it generates.
Furthermore, it can provide
user untraceability by means of the TOR network~\citep{tor} via the \emph{Orbot} application~\citep{orbot}.

A ChatSecure user may define several IM accounts (corresponding to one or more IM providers),
and use them at the same time to communicate with a set of \emph{buddies} (i.e., other
IM accounts (s)he is in contact with).

ChatSecure provides the typical functionalities of all IM applications, namely:
(a) contact management (i.e., inviting and removing contacts, accepting or 
denying invitations, etc.), (b) point-to-point communication, (c) group chats 
creation and participation, and (d) file transfer, as well as additional 
functionalities related to security management (toggling OTR encryption on and off, 
verification of the partner identity, etc.).

In this section we provide a detailed forensic analysis of ChatSecure
that is aimed at identifying all the relevant artifacts it generates,
interpreting them, and  using them to reconstruct the activities
carried out by its users.

In particular, after describing a fictitious scenario to
give an investigative context to the analysis techniques described in this paper,
we describe the set of artifacts generated
by ChatSecure, and where they are stored on the memory of the device
(Sec.~\ref{artifacts}).
Next, we illustrate how to reconstruct the set of ChatSecure accounts
utilized by the user (Sec.~\ref{accounts}), as well as 
the set of the corresponding buddies (Sec.~\ref{contacts}).
Then, we move to the reconstruction of the chronology and contents of 
exchanged messages (Sec.~\ref{chat-reconstruction}) and
files (Sec.~\ref{media}).
After that, we deal with the problem of decrypting the SQLite databases
where ChatSecure artifacts are stored (Sec.~\ref{encryption}) by describing
both a decryption algorithm we devised, and how the passphrase allowing
decryption can be retrieved from the volatile memory of the device.
Finally, we conclude with Sec.~\ref{sec:deletion}, where we 
report our findings concerning the deletion of the data generated by ChatSecure,
that show the impossibility of recovering them after they have been deleted.
\subsection{Investigative scenario}
\label{scenario}
To illustrate how the techniques discussed in this paper can be applied
in the context of a digital investigation, we consider the fictitious investigative
scenario reported below.

We consider the case where ChatSecure is found to be installed on a seized Android 
smartphone, and we need to answer the following set of typical investigative questions
(for each one of them, we also indicate the section of the paper where we discuss how
to obtain the corresponding answer):
\begin{enumerate}
	\item How many distinct XMPP accounts did the user configure and use
	with ChatSecure? (Sec.~\ref{accounts})
	\item Who are the ChatSecure contacts of the local user? (Sec.~\ref{contacts})
	\item What messages have been exchanged with each one of the above contacts, and when did each communication occur? (Sec.~\ref{chat-reconstruction})
	\item Did the local user exchange any file with its contacts? If so, when did these exchanges occur?
	What is the content of the files that have been exchanged? (Sec.~\ref{media})
	\item How to decrypt ChatSecure databases?  (Sec.~\ref{encryption})
	\item How to recover deleted data? (Sec.~\ref{sec:deletion}).
\end{enumerate}

\subsection{Location and format of ChatSecure artifacts}
\label{artifacts}
\begin{figure}[hbt]
	\centering
	\includegraphics[scale=0.5]{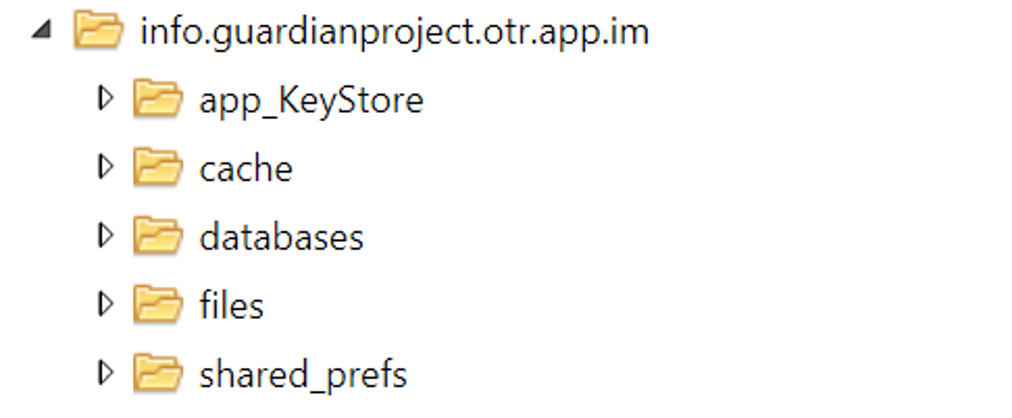}
	\caption{Structure of the main folder of ChatSecure.}
	\label{folders}
\end{figure}

During its use, ChatSecure stores several artifacts into various files and databases 
that are located into the  \linebreak\textit{info.guardianproject.otr.app.im} folder.
This folder is located into the \linebreak\textit{/data/data} directory of
the Android file system,~\footnote{This directory corresponds to the 
\emph{user data} partition of the internal device memory and is
inaccessible to standard users, unless the smartphone has been rooted.}
that contains the sub-folders shown in Fig.~\ref{folders}. 

The data of forensic interest generated by ChatSecure are the
following ones:
\begin{itemize}
	\item The \emph{main database}, where ChatSecure stores the information concerning
	the accounts used by the ChatSecure, the list of the corresponding buddies,
	and local copies of the messages that have been exchanged.
	It consists in an SQLCipher-encrypted SQLite v.3 database, named \textit{impsenc.db},
	which is stored in the \textit{databases} folder and contains
	21 different tables.
	As resulting from our findings, only 11 out of these 21 tables contain information of forensic interest, 
	namely tables \textit{accounts}, \textit{accountStatus}, \textit{providers}, 
	\textit{providerSettings}, \textit{contacts}, \textit{contactList}, \textit{presence},
	\textit{avatars}, \textit{chats}, \textit{messages}, and \textit{inMemoryMessages}.
	The information stored in these tables, as well their structure, interpretation,
	and analysis, are discussed in Secs.~\ref{accounts}--\ref{chat-reconstruction}, where we show how to
	use the data they store to perform various forensic reconstructions.
	The analysis of these tables has been performed by inspecting the source
	code of ChatSecure,~\footnote{In particular, files \textit{Imps.java} and \textit{ImpsProvider.java}, 
		stored in the \textit{src/info/guardianproject/otr/app/im/provider} directory of
		the ChatSecure source tree.} and by performing a set of controlled experiments in
	order to validate our findings.
	It is worth pointing out that the user may opt for not using encryption;
	this decision must be taken when ChatSecure is started for the 
	first time after installation, and cannot be undone (in this case
	the database is named \textit{imps.db}).
	\item The \emph{encrypted virtual disk}: in addition to exchanged messages,
	ChatSecure stores locally also copies of the files that its user has exchanged
	with her contacts. To prevent unauthorized parties from accessing these files, ChatSecure
	stores them into an encrypted virtual disk, which is implemented via IOCipher.
	The analysis of this encrypted virtual disk is discussed in Sec.~\ref{media}.
	\item The \emph{stored secret file}: 
	ChatSecure stores the information it needs to decrypt the main database and the
	virtual disk into
	a file, named \linebreak\textit{info.guardianproject.cacheword.prefs.xml}, which is located
	in folder \textit{shared\_prefs}. In Sec.~\ref{encryption} we show how this information
	can be decoded and used to carry out the above decryption.
\end{itemize}
For the sake of completeness, we mention also files \textit{account.xml} (storing
the information concerning the ChatSecure account) and \linebreak\textit{info.guardianproject.otr.app.im\_preferences.xml}
(storing ChatSecure settings and preferences), located in the
\textit{shared\_prefs} folder.
\subsection{Reconstructing user accounts}
\label{accounts}
As mentioned before, ChatSecure allows its users to create various IM 
accounts, each one corresponding to a specific IM provider.
From the investigative point of view, the information about all the
active accounts is relevant for various reasons, including determining
the identity of the providers to which additional sources of evidence
(e.g., log files) can be asked, and correlating evidence
with that retrieved from the devices of other ChatSecure users with whom
the user has exchanged communications.

The information associated with each account (name, credentials, etc.) are 
stored in the main database, where they are spread across four distinct tables
(namely, \textit{accounts}, \textit{accountStatus}, \textit{providers}, and 
\textit{providerSettings}).
The structure of these tables and the meaning of their fields are reported
in Tables~\ref{tab:accounts}--\ref{tab:providerSettings}.

Tables \textit{accounts} and \textit{accountStatus} jointly store the information
concerning the IM accounts created by the ChatSecure user.
In particular, \textit{accounts} stores the properties of these accounts
(e.g., the credentials for the authentication), while \textit{accountStatus} stores
the information concerning their \emph{status}.
These two tables are linked together by means of the foreign key of table
\emph{accountStatus}  (i.e., field \textit{account}, see Table~\ref{tab:accountStatus}).

\begin{table*}[hbp]
	\caption{Structure of the \textit{accounts} table.}
	\label{tab:accounts}
	\begin{center}
		\begin{footnotesize}
			\begin{tabularx}{\linewidth}{|l|l|l|X|}
				\hline
				\multicolumn{4}{|c|}{\textbf{Table \textit{accounts}}} \\ \hline
				\emph{Name} & \emph{Role} &\emph{Type} & \emph{Meaning} \\ \hline
				\_id & Primary Key & int & unique record identifier 
				\\ \hline
				name & -- & text & name of the account as chosen by the user\\ 
				\hline
				provider & Foreign Key & int & value of the \textit{\_id } field of the record, in table \textit{providers}, 
				corresponding to the provider of this IM account\\ 
				\hline
				username & -- & text & username (on the service provider) for the 
				account\\ \hline
				pw & -- & text & password (on the service provider) for the 
				account \\ \hline
				active & -- & int & $1$ if the account is active, $0$ otherwise \\ 
				\hline
				locked & -- & int & $1$ if the account is locked (i.e., it is not 
				editable), $0$ otherwise \\ \hline
				keep\_signed\_in & -- &  int & $1$ if ChatSecure keeps the account 
				logged in between executions, $0$ otherwise \\ \hline
				last\_login\_state & -- & int & either $0$ or $1$ \\ \hline
			\end{tabularx}
		\end{footnotesize}
	\end{center}
\end{table*}

\begin{table*}[hbtp]
		\caption{Structure of the \textit{accountStatus} table.}
		\label{tab:accountStatus}
	\begin{center}
		\begin{footnotesize}
			\begin{tabularx}{\linewidth}{|l|l|l|X|}
				\hline
				\multicolumn{4}{|c|}{\textbf{Table \textit{accountStatus}}} \\ \hline
				\emph{Name} & \emph{Role } & \emph{Type} & \emph{Meaning} \\ \hline
				\_id & Primary Key & int & unique record identifier  
				\\ \hline
				account & Foreign Key & int & value of the \textit{\_id} field of the record, in the \textit{accounts} table,
				corresponding to the account this status information refers to \\ \hline
				presenceStatus & -- & int & visibility of the account to its 
				buddies. Possible values are: $0$ (offline), $1$ 
				(invisible), $2$ (away), $3$ (idle), $4$ (do not disturb), 
				and $5$ (available) \\ \hline
				connStatus & -- & int & status of the connection to the XMPP 
				provider. Possible values are: $0$ (offline),
				$1$ (connecting), $2$ (suspended due to temporary network 
				unavailability), and $3$ (online) \\ \hline
			\end{tabularx}
		\end{footnotesize}
	\end{center}

\end{table*}

Tables \textit{providers} and \textit{providerSettings}, instead, jointly
store the information about the IM providers corresponding to the above IM accounts.
In particular, the former table stores the information about these providers,
while the latter one stores the settings of each provider (one record per setting).
These two tables are joined together 
by means of the foreign key of table \emph{providerSettings}  (i.e., field \textit{provider}, 
see Table~\ref{tab:providerSettings}).

\begin{table*}[hbtp]
	\caption{Structure of the \textit{providers} table (fields \textit{category} and \textit{signup\_url} 
		have been omitted because of their lack of forensic value).}
	\label{tab:providers}
	\begin{center}
		\begin{footnotesize}
			\begin{tabularx}{\linewidth}{|l|l|l|X|}
				\hline
				\multicolumn{4}{|c|}{\textbf{Table \textit{providers}}} \\ \hline
				\emph{Name} & \emph{Role} & \emph{Type} & \emph{Meaning} \\ \hline
				\_id & Primary Key & int & unique record identifier \\ \hline
				name & -- & text & name of the IM provider (e.g.\ GTalk, AIM, 
				etc.)\\ \hline
				fullname & -- &  text & full name of the IM provider\\ \hline
			\end{tabularx}
		\end{footnotesize}
	\end{center}
	
\end{table*}

\begin{table*}[hbtp]
	\caption{Structure of the \textit{providerSettings} table.}
	\label{tab:providerSettings}
	\begin{center}
		\begin{footnotesize}
			\begin{tabularx}{\linewidth}{|l|l|l|X|}
				\hline
				\multicolumn{4}{|c|}{\textbf{Table \textit{providerSettings}}} \\ \hline
				\emph{Name} & \emph{Role} & \emph{Type} & \emph{Meaning} \\ \hline
				\_id & Primary Key & int &  unique record identifier \\ \hline
				provider & Foreign Key & int & value of the \textit{\_id} field of the record, in the \emph{providers} table,
				corresponding to the IM provider this setting refers to \\ \hline
				name & -- & text & name of the setting\\ \hline
				value & -- & text & value of the setting \\ \hline
			\end{tabularx}
		\end{footnotesize}
	\end{center}	
\end{table*}

\begin{figure}[hbt]
	\centering
	\includegraphics[scale=0.70]{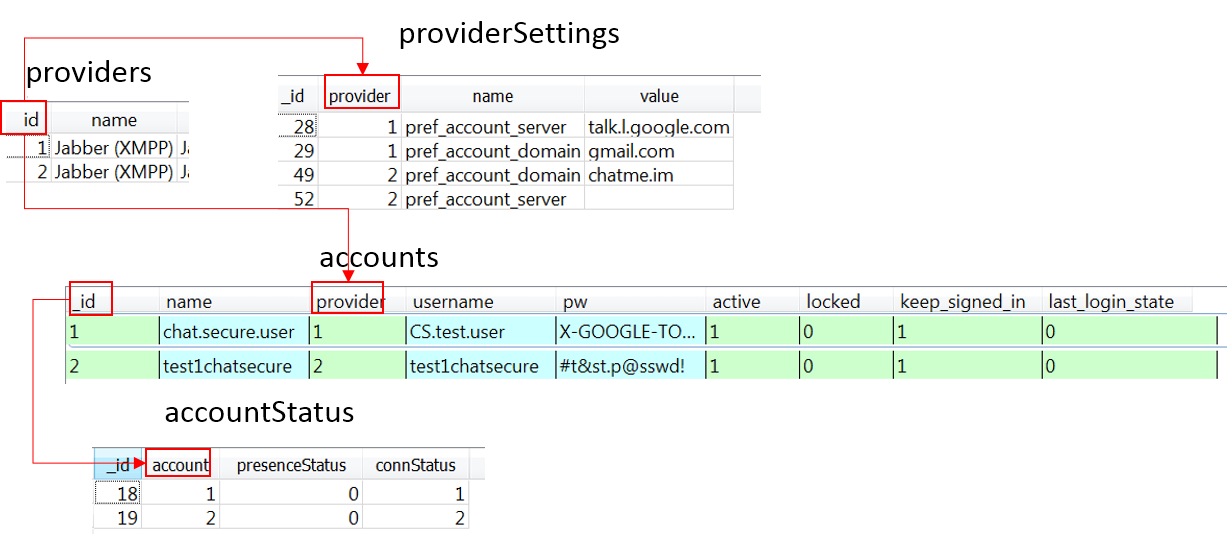}
	\caption{Reconstruction of ChatSecure user accounts.}
	\label{account-tables}
\end{figure}

To illustrate how to reconstruct the information concerning the ChatSecure accounts, 
let us consider the scenario shown in Fig.~\ref{account-tables}.

This scenario features two distinct ChatSecure accounts, 
named \linebreak\textit{chat.secure.user} and \textit{test1chatsecure}
(see the records stored in table \textit{accounts} in Fig.~\ref{account-tables}).
Both these accounts are \emph{active} (i.e., they are currently used), as indicated
by the value $1$ stored in fields \textit{active}, and are never logged out of the respective IM provider
during a ChatSecure session, as indicated by the value $1$ stored in fields \textit{keep\_signed\_in}.

To determine the IM provider associated with each account we need to join
tables \textit{accounts} and \textit{providers}.
The results of this operation show that user \textit{chat.secure.user} is associated with IM 
provider no.\ 1 and that authenticates with it using username \textit{CS.test.user}
and password ``\textit{X-GOOGLE-TOKEN\dots}'' (password shortened for readability purposes),
while user \linebreak\textit{test1chatsecure} is associated with IM
provider no.\ 2 and authenticates with it using username \textit{test1chatsecure}
and password ``\textit{\#t\&st.p@sswd!}''.
Note that the passwords used to authenticate
with IM providers are stored in cleartext, and as such may be readily
used to authenticate with the IM providers \emph{outside the ChatSecure application} 
once recovered from this database.

To determine the identity of the IM providers used by the various accounts
we need to join tables \textit{providers} and \textit{providerSettings}.
The results of this operation indicate that provider no.\ 1 corresponds to Google's \textit{GTalk} IM service
(see record no.\ 28 of table \textit{providerSettings} in Fig.~\ref{account-tables}), and that its account
domain is \textit{gmail.com} (see record no.\ 29 of table \textit{providerSettings}).
From this, we determine that the ChatSecure user \textit{chat.secure.user} corresponds to the GTalk 
user \textit{CS.test.user@gmail.com}.
Furthermore, we also determine that provider no.\ 2 corresponds to
the \textit{ChatMe} IM service (it uses the \textit{chatme.im} server),
and that its account domain is \textit{chatme.im}, meaning that
ChatSecure user \textit{test1chatsecure} corresponds to
ChatMe user \linebreak\textit{test1chatsecure@chatme.im}.

Finally, to determine the status of each account we need to join
tables \textit{accounts} and \textit{accountStatus}.
From the results of this operation we determine that
both users \textit{chat.secure.user} and \textit{test1chatsecure} 
are  offline because of the unavailability of the network when
the memory of the device was acquired (see the values stored in fields \textit{presenceStatus} and 
\textit{connStatus} in the records of table \textit{accountStatus}).
\subsection{Reconstructing contact lists}
\label{contacts}
Each ChatSecure account is typically associated with a set of contacts, i.e.
remote users with whom (s)he can exchange messages and files.
The evidentiary value of contact information is notorious, as it allows an investigator
to determine who the user was in contact with.

Each one of the contacts is associated with its \emph{nickname} (i.e., the name used
by the ChatSecure user to denote the buddy),
its \emph{username} (that identifies the contact on the corresponding IM provider),
and an optional \emph{avatar} (i.e., a picture, chosen by the corresponding 
user, which is downloaded by ChatSecure and displayed together with the nickname).

The information concerning user contacts is stored in the main database,
and is spread across four distinct tables, namely \textit{contacts}, \textit{avatars}, 
\textit{presence}, and \textit{contactsList}, whose structure and meaning is
reported in Tables~\ref{tab:contacts}--\ref{tab:contactList}.

\begin{table}[hbtp]
	\caption{Structure of the \textit{contacts} table (fields \textit{qc} and \textit{rejected} 
		have been omitted because of their lack of forensic value).}
	\label{tab:contacts}
	\begin{center}
		\begin{footnotesize}
			\begin{tabularx}{\linewidth}{|l|l|l|X|}
				\hline
				\multicolumn{4}{|c|}{\textbf{Table \emph{contacts}}} \\ 
				\hline
				\emph{Name} & \emph{Role} & \emph{Type} & \emph{Meaning} \\ \hline
				\_id	& Primary Key & int &  unique identifier of the contact \\ \hline
				username     & Secondary Key & text & username of this contact on the 
				corresponding IM provider\\ \hline
				nickname  & -- & text & name displayed by ChatSecure for this 
				contact \\ \hline
				provider	& Foreign Key & int & value of field \textit{\_id} of the record, in table \textit{providers},
				this contact is an account of \\ \hline
				account		& Foreign Key & int & value of field \textit{\_id} of the record, in table \textit{accounts},
				corresponding to the local ChatSecure account this contact belongs to \\ \hline
				contactList	& Foreign Key & int & value of field \textit{\_id} of the record, in table \textit{contactList},
				corresponding to the contact list this contact belongs to \\ \hline
				type & -- & int & contact type: $0$ (normal), $1$ (temporary, not in contacts but 
				subscribed to receive updates), $2$ (temporary group chat contact), $3$ (blocked), $4$ (hidden), $5$ (pinned)\\ \hline
				subscriptionStatus	& -- & int & status update receipt from this contact: $0$ 
				(none), $1$ (requested to subscribe), $2$ (requested to 
				unsubscribe) \\ \hline
				subscriptionType	& -- & int & exchange of status updates
				with this contact: $0$ (no 
				interest in update), $1$ (stop receiving updates), $2$ (receive updates), $3$ (contact wants updates from the 
				user), $4$ (mutual interest in receiving updates), $5$ (pending invitations)\\ \hline
				otr & -- & int & status of the OTR encryption; possible values are: 
				$0$ (off), $1$ (on, don't know who turned it on), 
				$2$ (on, enabled by the user), $3$ (on, enabled by the contact) \\ 
				\hline 
			\end{tabularx} 
		\end{footnotesize}
	\end{center}
\end{table}

\begin{table}[hbtp]
	\caption{Structure of table \textit{presence} (fields \textit{jid\_resource} and \textit{priority} 
		have been omitted because of their lack of forensic value).}
	\label{tab:presence}
	\begin{center}
		\begin{footnotesize}
			\begin{tabularx}{\linewidth}{|l|l|l|X|}
				\hline
				\multicolumn{4}{|c|}{\textbf{Table \textit{presence}}} \\ \hline
				\emph{Name} & \emph{Role} & \emph{Type} & \emph{Meaning} \\ \hline
				\_id & Primary Key & int &  unique identifier of the record \\ \hline
				
				contact\_id& Foreign Key & int & value of field \textit{\_id} of the record, in 
				table \textit{contacts} table, corresponding to the contact this presence information refers to \\ \hline
				client\_type & -- & int & type of the client; possible values are: $0$ (default), $1$ (mobile), $2$ (android)\\ \hline
				mode & -- & int & presence status of the contact; possible values are: $0$ (offline), $1$ (invisible), $2$ (away), $3$ (idle), $4$ (do not 
				disturb), and $5$ (available)  \\ \hline
				status & -- & text & status message of the contact \\ \hline
			\end{tabularx}
		\end{footnotesize}
	\end{center}
	
\end{table}
\begin{table}[hbtp]
	\caption{Structure of the \textit{avatars} table.}
	\label{tab:avatars}
	\begin{center}
		\begin{footnotesize}
			\begin{tabularx}{\linewidth}{|l|l|l|X|}
				\hline
				\multicolumn{4}{|c|}{\textbf{Table \textit{avatars}}} \\ \hline
				\emph{Name} & \emph{Role} & \emph{Type} & \emph{Meaning} \\ \hline
				\_id & Primary key & int &  unique avatar identifier\\ \hline
				contact & Foreign Key & text & value of field \textit{username} of the record, in table \textit{contacts}, this avatar belongs to \\ \hline
				provider & Foreign Key & int & value of field \textit{\_id} of the record, in table \textit{providers}, corresponding to the provider this 
				contact is user of \\ \hline
				account & Foreign Key & int & value of field \textit{\_id} of the record, in table  \textit{accounts}, corresponding to the account the contact 
				owning this avatar is buddy of \\ \hline
				hash & -- & text & SHA-1 hash of the picture used as avatar by this contact \\ \hline
				data & -- & blob & raw image data of the avatar \\ \hline
			\end{tabularx}
		\end{footnotesize}
	\end{center}
	
\end{table}

\begin{table}[hbtp]
	\caption{Structure of table \textit{contactList}.}
	\label{tab:contactList}
	\begin{center}
		\begin{footnotesize}
			\begin{tabularx}{\linewidth}{|l|l|l|X|}
				\hline
				\multicolumn{4}{|c|}{\textbf{Table \textit{contactList}}} \\ \hline
				\emph{Name} & \emph{Role} & \emph{Type} & \emph{Meaning} \\ \hline
				\_id & Primary Key & int &  unique identifier of the record\\ \hline
				name & -- & text & display name of this contact list\\ \hline
				account & Foreign Key & int & value of field \textit{\_id} of the record, in table \textit{accounts},
				corresponding to the ChatSecure account this list belongs to \\ 
				\hline
				provider & Foreign Key & int & value of field \textit{\_id} of the record, in table  
				\textit{providers table}, corresponding to the provider of 
				the ChatSecure account this contact list belongs to \\ \hline
			\end{tabularx}
		\end{footnotesize}
	\end{center}
	
\end{table}

In particular:
\begin{itemize}
	\item tables \textit{contacts} and \textit{presence} store the information about the 
	various contacts and on the corresponding status, respectively, and are linked
	together via the foreign key of table \textit{presence} (i.e., field \textit{contact\_id},
	see Table~\ref{tab:presence});
	\item table \textit{avatars} stores the information about the 
	avatars of the various contacts and, as reported in Table~\ref{tab:avatars},
	it is linked to various tables,
	namely \textit{contacts}, \textit{providers}, and \textit{accounts},
	via the corresponding foreign keys;
	\item table \textit{contactList} stores the information about how
	the contacts of the ChatSecure users are organized into lists, and is linked
	to both tables \textit{accounts} and \textit{provider} (to associate the list with the 
	specific ChatSecure user the list belongs to) via the corresponding 
	foreign keys (see Table~\ref{tab:contactList}).
\end{itemize}

\begin{figure}[hbtp]
	\centering
	\includegraphics[scale=0.6]{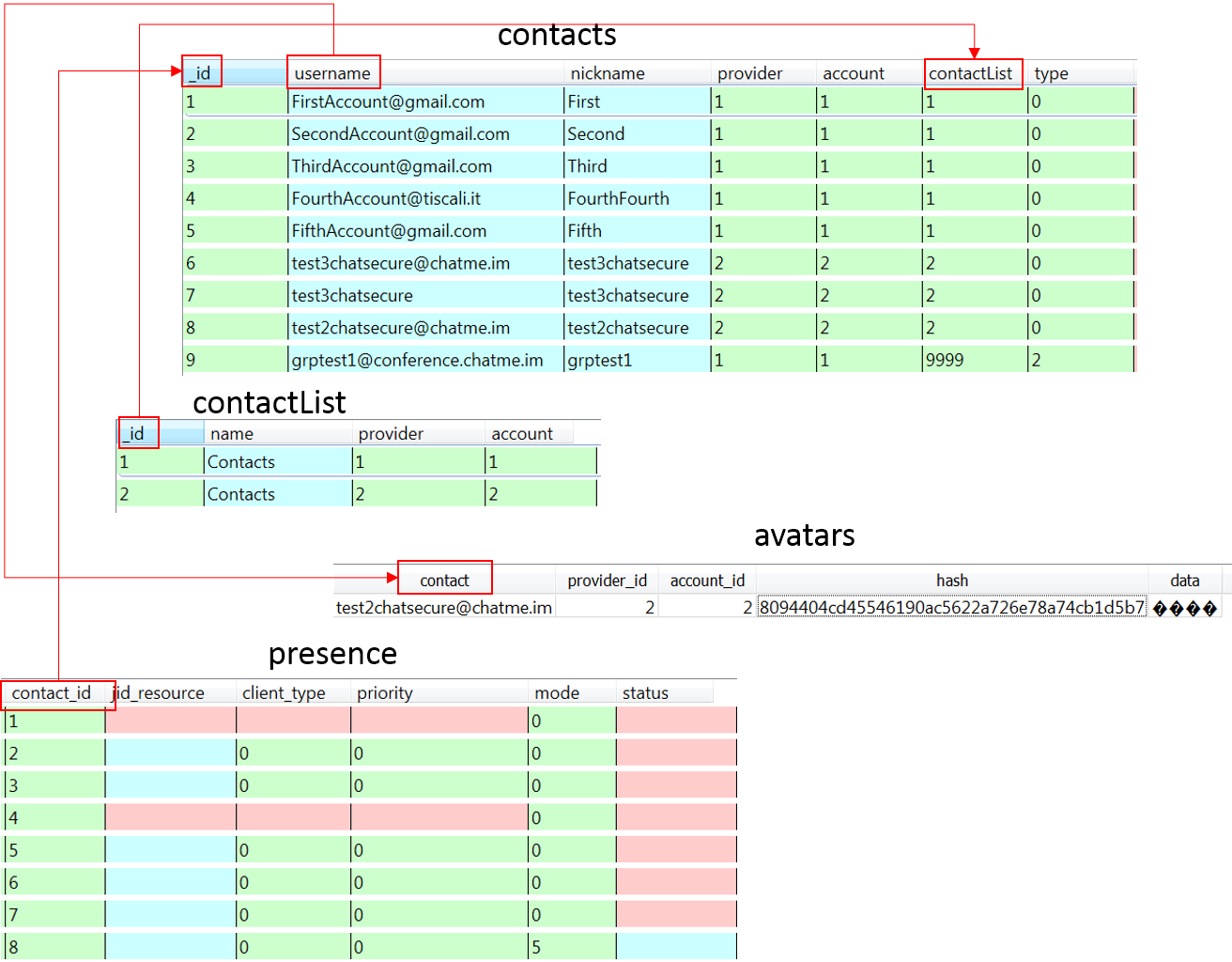}
	\caption{Reconstruction of ChatSecure contact lists.}
	\label{contact-tables}
\end{figure}

To illustrate how to reconstruct the information about the contacts of
a ChatSecure user, let us consider the
scenario shown in Fig.~\ref{contact-tables}, where 9 distinct 
contacts belonging to the two ChatSecure accounts shown in Fig.~\ref{account-tables}  
(namely, \textit{chat.secure.user} and \textit{test1chatsecure})
are stored in table \textit{contacts}.

From records no.\ $1,\ldots,5$ and $9$, we see that six contacts
are associated with user account no.\ 1 (\textit{chat.secure.user}) and 
provider no.\ 1 (\textit{Google GTalk}), while the remaining three 
contacts (corresponding to records no.\ $6,7$, and $8$) are associated with user account 
no.\ 2 (\textit{test1chatsecure}) and provider no.\ 2 (\textit{ChatMe}).
Furthermore, we also see that all these contacts are of the ``normal'' type 
(field \textit{type}=$0$), with the exception of contact no.\ 9 
(\textit{nickname}=`\emph{grptest1}') that is a temporary contact created 
purposely to denote a group chat (\textit{type}=$2$, see Table~\ref{tab:contacts}) corresponding to
the remote IM account \textit{grptest1@conference.chatme.im}.

Avatar pictures may have evidentiary value as well: they can be indeed used to link
a ChatSecure contact to the real identity of the person using it (for instance, if
the avatar displays the face of the user, or any location or item that can be
uniquely associated with that person).

To determine the avatars associated with each contact, we join tables \textit{contacts} and \textit{avatars}.
The results of this operation indicate that
only contact no.\ 8 (user \emph{test2chatsecure@chatme.im}) is associated with an avatar, whose picture is stored 
(as a ``blob'' of bytes) in field \textit{data}; the avatar can be extracted from this field,
and visualized using a standard image viewer.

Also the status information of a contact may have evidentiary value.
For instance, the textual status message may provide information about the
real identity of the contact, and also its presence status at the moment
of the last update may provide information about the behaviour of that contact.

To determine the status of each contact, as reported by the last time this information
was updated locally, we have to join tables \textit{contacts} and \textit{presence}.
From the results of this operation, we see that (a) group chat contacts (i.e., contact no.\ $9$ in our example scenario) 
have no associated status, (b) the status all the other contacts is
``\textit{offline}'' (the value $0$ is stored in the corresponding \textit{mode} field), with the
exception of contact \textit{test2chatsecure@chatme.im} (contact no.\ $8$), whose
status is instead ``\textit{available}'' (\textit{mode} field contains $5$).
We also observe that none of these contacts is associated to a textual status message
(fields \textit{status} are empty).

Finally, also contact lists may be important from the evidentiary point of view, as they
allow to link each contact to the corresponding ChatSecure account used by the local
user.

To reconstruct the contact lists, we have to join tables \textit{contacts} and \textit{contactList}.
From the results of this operation, we see that
these contacts are organized in two distinct lists: the first one includes contacts
no.\ $1,\ldots,5$ and $9$, and belongs to the ChatSecure account no.\ 1, while the second one includes
the remaining contacts, and belongs to account no.\ 2. In both cases, the name of the list
is \textit{Contacts}.

\subsection{Reconstructing the chronology and contents of chat messages}
\label{chat-reconstruction}
Reconstructing the time in which each message was sent or received, the content
of that message, and the communication partner, is of obvious investigative importance.

Each time a message is sent or received, ChatSecure stores in the main database
a record containing both its textual content and various metadata
(e.g., the identifier of the corresponding buddy, and the date and time when the
exchange occurred).
This information  is spread across two 
distinct tables of the main database, namely \textit{messages} and 
\textit{inMemoryMessages}, that have the same structure,~\footnote{The main database 
	includes also another chat-related table, named \textit{chats},
	that has not been described here since the information it stores is redundant being it repeated
	in tables \textit{messages} and \textit{inMemoryMessages}.}
that is described in Table~\ref{tab:messages} together with the interpretation of its fields.
The reason for which two distinct tables are used is unclear; however, the messages they 
contain do not overlap, so both of them 
needs to be analyzed to recover all the messages that have been exchanged.
\begin{table*}[hbtp]
	\caption{Structure of the \textit{inMemoryMessage} and \textit{messages} tables (fields \textit{packet\_id} and
		\textit{shown\_ts} have been omitted because of their lack of forensic value).}
	\label{tab:messages}
	\begin{center}
		\begin{footnotesize}
			\begin{tabularx}{\linewidth}{|l|l|X|}
				\hline
				\multicolumn{3}{|c|}{\textbf{The  
						\textit{inMemoryMessages} and \textit{messages} 
						tables}} \\ \hline
				\emph{Name} & \emph{Type} & \emph{Meaning} \\ \hline
				\_id & int &  unique identifier of the message \\ \hline
				thread\_id & int & identifier of the contact this message has been exchanged with \\ \hline
				nickname & text & used for group chat messages only to indicate the nickname chosen by the local 
				ChatSecure user in that group chat (empty for one-to-one messages)\\ \hline
				body & text & body of the message \\ \hline
				date & int &  the date this message has been sent or received (13-digits Unix epoch format)\\ \hline
				type & int & type of the message: 
				$0$ (outgoing), $1$ (incoming), $2$ (presence became 
				``available''), $3$ (presence became ``away''), $4$ (presence 
				became ``busy''), $5$ (presence became ``unavailable''), $6$ 
				(message converted to a group chat), $7$ (status message), $8$ 
				(message cannot be sent now, will be sent later), $9$ (OTR is 
				turned off), $10$ (OTR is turned on), $11$ (OTR turned on by 
				the user), $12$ (OTR turned on by the communicating partner), 
				$13$ (incoming encrypted), $14$ (incoming encrypted and verified), $15$ 
				(outgoing encrypted), $16$ (outgoing encrypted and verified) \\ \hline
				err\_code & int & error code ($0$ = no error)\\ \hline
				err\_msg & text & error message (if any)\\ \hline
				is\_muc & int & flag indicating whether it is a group chat message ($1$) or not ($0$)\\ \hline
				is\_delivered & int & flag indicating whether a ``delivered'' confirmation was received ($1$) or not ($0$)\\ \hline
				mime\_type & text & type of data exchanged (null for text message, non-null for transferred files (see Sec.~\ref{media})\\ \hline
			\end{tabularx}
		\end{footnotesize}
	\end{center}
\end{table*}

From the analysis of the meaning of the various fields (and, in particular, of the possible values
of field \textit{type}), we see that ChatSecure messages belong to two distinct
categories, namely:
\begin{itemize}
	\item \emph{notification messages}, 
	i.e.\ messages that do not carry any user-generated content, but that
	instead carry updates about the status contact, such as changes of
	his/her status (message types $2, 3, 4$ and $5$),
	or of his/her OTR encryption status (message types $9, 10, 11$ and $12$).
	\item \emph{chat messages}, that carry user-generated textual content.
	These messages correspond to records whose \textit{type} fields
	stores values in the set $\{0,1,13,14,15\}$ (see Table~\ref{tab:messages})
	to denote cleartext outgoing (\emph{type}=$0$) or incoming (\emph{type}=$1$) messages, encrypted
	incoming message sent by an unverified (\emph{type}=$13$) or a verified (\emph{type}=$14$) partner,
	and encrypted outgoing messages sent to an unverified (\emph{type}=$15$) or verified
	(\emph{type}=$16$) partner.
\end{itemize}

The chronology of message exchanges can be reconstructed by means of the values
stored in the \textit{date} field, that store date and time of message transmission or receipt
encoded as a 13-digits Unix epoch format. This holds true both for notification and chat messages,
so it is possible to reconstruct not only the chronology of messages exchanged by users,
but also when a notification message arrived.

\begin{figure}[hbtp]
	\centering
	\includegraphics[scale=0.6]{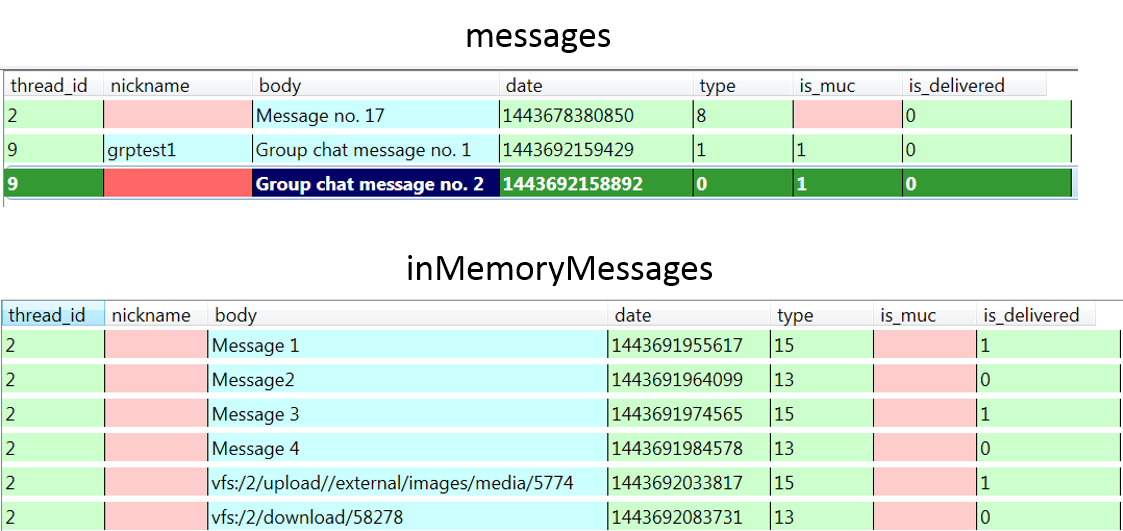}
	\caption{Reconstruction of chronology and content of exchanged messages.}
	\label{messages-tables}
\end{figure}

To illustrate how to reconstruct the chronology of exchanged messages and
the corresponding contents, let us consider the scenario depicted in 
Fig.~\ref{messages-tables} that shows 10 exchanged messages, which are stored 
in Tables \textit{messages} (4 messages) and \textit{inMemoryMessages} (6 messages).

From this figure, we see that the first record in table \textit{inMemoryMessage}, 
corresponds to an outgoing encrypted message (\textit{type}=$15$) that was sent
on Oct.\ $1^\text{st}$, 2015 at 9:32:35.617 a.m.\ UTC (\textit{date}=`\textit{1443691955617}')
to contact no.\ $2$ (\textit{thread\_id}=$2$) (that corresponds to the contact whose nickname is \emph{Second},
see Fig.~\ref{contact-tables}); the message body was 
``\textit{Message 1}'', and has been successfully delivered to (i.e., visualized by)
its recipient (\textit{is\_delivered}=$1$).
The second record of \textit{inMemoryMessages} correspond instead to an incoming encrypted message
(\textit{type}=$13$), that was received by the same user and
carried as textual content the string ``\textit{Message 2}'' on Oct.\ $1^\text{st}$, 2015
at 9:32:44.099 a.m.\ UTC (\textit{date}=`\textit{1443691964099}') that has not been
delivered (i.e., visualized) by the ChatSecure user.

Finally, the first message of table \textit{messages} (whose body was 
``\textit{Message no.\ 17}''), is an outgoing message
sent to the same contact on Oct.\ $1^\text{st}$, 2015 
at 05:46:20.850 a.m.\ UTC (\textit{date}=`\textit{1443678380850}'),
but whose transmission was delayed (\textit{type}=$8$) and, as such, had not been 
successfully delivered to the recipient (\textit{is\_delivered}=$0$).

Note that ChatSecure stores in tables \textit{messages} and \textit{inMemoryMessages} 
the messages corresponding to all
the local accounts, i.e.\ \textit{chat.secure.user} and \textit{test1chatsecure}
in our example (see Fig.~\ref{account-tables}).
However, the identity of the local account $LA$ corresponding to a given message can be
easily determined by correlating the unique contact identifier $ID$ stored in the 
\textit{thread\_id} field with the record of table \textit{contact} storing
$ID$ in its \textit{\_id} field; the value of field \textit{account} of that
record will indicate the local account $LA$.
In this way, in the examples above we could tell that messages have been exchanged
between contact no.\ $2$ and account \textit{chat.secure.user}, since this contact
is associated with account no.\ $1$ (see Fig.~\ref{account-tables}).

Table \textit{messages} in Fig.~\ref{messages-tables} also stores messages
exchanged in group chats, corresponding to the last two records of table \textit{messages},
as indicated by \textit{is\_muc=1}. From these records, we see that the corresponding messages
have been sent to the group chat corresponding to the contact no.\ \textit{9} of
table \textit{contacts} (the chat room named \textit{grptest@conference.chatme.im}),
and carried as textual content the strings ``\textit{Group chat message no.\ 1}''
and ``\textit{Group chat message no.\ 2}'', respectively.

As a final observation, it is worth noticing that ChatSecure stores messages in cleartext, 
even if they have been encrypted before transmission
(see the contents of field \textit{body} of all the
records in Fig.~\ref{messages-tables}): this is indeed the case of all the records stored in
table \textit{inMemoryMessages}, that correspond to encrypted messages that 
have been either sent (\textit{type}=$15$) or received (\textit{type}=$13$).

\subsection{Reconstructing the chronology and contents of file exchanges}
\label{media}
In addition to textual messages,
ChatSecure allows its users to exchange also files of any type
(at the moment of this writing, however, this functionality 
is available only for one-to-one communications and not for group chats).
Determining the chronology of these exchanges, and more importantly the contents
of exchanged files, may be of crucial importance in many investigations.

Each time a file is exchanged, ChatSecure creates a record that stores the same information described for
chat messages,
either in the \textit{messages} or in the \textit{inMemoryMessages} of the
main database .
Furthermore, it stores the content of the file into
an IOCipher encrypted virtual disk
to keep it inaccessible to an authorized third-party.
The file transfer mechanism used by ChatSecure interfaces directly with its
encrypted virtual disk, that is: (a) before being sent, files are stored on
the virtual disk, from which they are fetched and sent across
the network, and (b) received files are stored directly in the virtual disk.

The message records corresponding to file transfers
are identified looking at the contents of their \textit{mime\_type} and
\textit{body} fields.
In particular, the former field stores the MIME media type~\citep{mime} of the transferred 
file, while the latter one stores the full path of the file
in the encrypted virtual disk.

IOCipher implements the above virtual disk by using 
\emph{libsqlfs}~\citep{libsqlfs}, a library that in turn 
implements a POSIX-style  file system by means of an SQLCipher-encrypted
SQLite database.
This database is named \textit{media.db}, and includes only two tables,
namely \textit{meta\_data} (storing various file metadata, such as identifier, path name,
and timestamps),
and \textit{value\_data} (storing the actual file blocks),
whose structure and interpretation is reported in Tables~\ref{media:metadata} and \ref{media:valuedata},
respectively.
\begin{table*}[hbtp]
	\caption{Structure of table \textit{meta\_data} of the \textit{media.db} database 
		(fields lacking any forensic value are omitted).}
	\label{media:metadata}
	\begin{center}
		\begin{footnotesize}
			\begin{tabularx}{\linewidth}{|l|l|X|}
				\hline
				\multicolumn{3}{|c|}{\textbf{Table   
						\textit{meta\_data}}} \\ \hline
				\emph{Name} & \emph{Type} & \emph{Meaning} \\ \hline
				type & text & type of the object: directory (\textit{dir}), file (\textit{blob}), symbolic link (\textit{symlink}) \\ \hline 
				key & text & full path of the object in the libsqlfs file system\\ \hline
				ctime, mtime, atime & int & file creation, last modification, and last access time, respectively (10-digits Unix epoch format) \\ \hline
				size & int & file size (in bytes)\\ \hline
				block\_size & int & block size (in bytes)\\ \hline		
			\end{tabularx}
		\end{footnotesize}
	\end{center}
\end{table*}
\begin{table*}[hbtp]
	\caption{Structure of table \textit{value\_data} of the \textit{media.db} database 
		(fields lacking any forensic value are omitted).}
	\label{media:valuedata}
	\begin{center}
		\begin{footnotesize}
			\begin{tabularx}{\linewidth}{|l|l|X|}
				\hline
				\multicolumn{3}{|c|}{\textbf{Table   
						\textit{value\_data}}} \\ \hline
					\emph{Name} & \emph{Type} & \emph{Meaning} \\ \hline
					key & text & full path of the file in the libsqlfs file system, as stored in the corresponding
					\textit{meta\_data} table\\ \hline
					block\_no & int & sequence number of the file block stored in this record\\ \hline 
					data\_block & binary & data stored in the file block corresponding to this record \\ \hline	
				\end{tabularx}
			\end{footnotesize}
		\end{center}
	\end{table*}

To reconstruct the chronology and the contents of the files that have been exchanged,
it is necessary to analyze and correlate the records stored both in the
main database \textit{impsenc.db} and the \textit{media.db} database implementing the encrypted virtual
disk.

\begin{figure}[hbtp]
	\centering
	\includegraphics[scale=0.6]{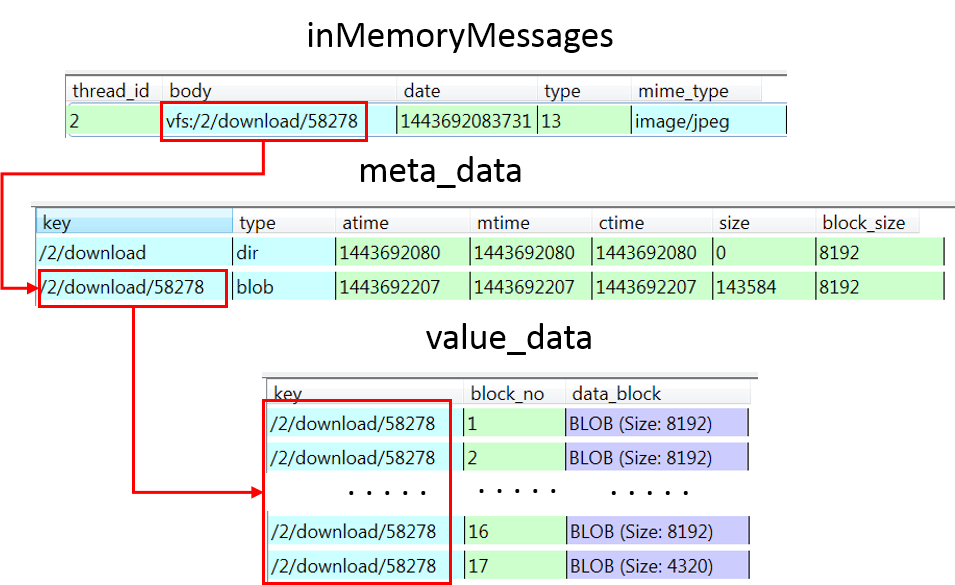}
	\caption{Reconstruction of a downloaded file.}
	\label{media-tables}
\end{figure}

To illustrate how to perform this reconstruction, 
let us consider Fig.~\ref{media-tables}, that shows the 
records generated during the download of a file (for the upload case, the 
situation is similar).

First, it is necessary to identify the records of tables \textit{messages} and
\textit{inMemoryMessages} corresponding to file transfers by examining the values
stored in the \textit{mime\_type} and \textit{body} fields.
Fig.~\ref{media-tables} shows the record corresponding to the transfer of a
JPEG image (\textit{mime\_type=`image/jpeg'}) which has been stored in the encrypted virtual
disk into a file whose full path name is `\textit{/2/download/58278}'
(the `\textit{vfs:}' prefix is not part of the path name,
but only indicates the use of the Android Virtual File System to interface
with the libsqlfs file system).
From this record, we can also determine that it was an incoming encrypted file 
(\textit{type=13}), received from contact no.\ \textit{2} on Oct.\ $1^\text{st}$, 2015 at 
09:34:43.731 a.m.\ UTC (\textit{date=`1443692083731'}).

After having identified the files that have been exchanged, it is possible
to retrieve the corresponding data by examining the records stored in the
tables of the \textit{media.db} database.
The starting point is table \textit{meta\_data}, in which we search
for a record whose \textit{key} field stores the same path name stored
in the \textit{body} field of the corresponding
\textit{inMemoryMessages} record, i.e.\ `\textit{/2/download/58278}'.
From this record, we can determine that the file of interest
has a size of $143584$ bytes, and is stored as a sequence of blocks of $8192$ bytes each.

To retrieve these blocks, 
all the records of table \textit{value\_data} whose \textit{key} field
stores the value `\textit{/2/download/58278}' must be retrieved, and then
the content of their \textit{data\_block} field must be extracted to be
stored into a single file
according to the corresponding sequence numbers.

\subsection{Dealing with encryption}
\label{encryption}
As mentioned before, ChatSecure relies on SQLCipher to encrypt,
using the AES-256 algorithm, 
both the main database \textit{impsenc.db}, as well as the
\textit{media.db} database used by IOCipher to implement the
encrypted virtual disk.
Therefore it is necessary to decrypt them in order to analyze their contents.

The encryption key used by SQLCipher is generated internally by ChatSecure,
and is never exposed to the user. This key is instead saved in the
internal memory of the device so that it can be retrieved and used
by ChatSecure to decrypt the above databases.
However, to make sure that an adversary cannot decrypt these databases using the
saved secret key, ChatSecure
uses the \emph{CacheWord}~\citep{cacheword} library
to encrypt it using a user-defined \emph{secret passphrase}, and to store it
into an XML file named \textit{info.guardianproject.cacheword.prefs.xml} located
in the \textit{shared\_prefs} folder (see Fig.~\ref{folders}).
To decrypt the saved secret key, the passphrase set by the user
needs to be re-entered each time ChatSecure is started.

From the above discussion it follows that to decrypt the ChatSecure databases 
three distinct problems must be solved, namely:
\begin{enumerate}
	\item obtaining the \emph{secret passphrase} chosen by the user;
	\item decrypting the \emph{secret key} stored by CacheWord;
	\item decrypting the databases using the secret key.
\end{enumerate}

In the rest of this section, after describing the encryption scheme adopted
by ChatSecure (Sec.~\ref{sec:encryption}), we discuss how to decrypt the secret
database encryption key (Sec.~\ref{sec:decrypt-key}) and how to decrypt the 
databases using this key (Sec.~\ref{sec:decrypt-db}).
Finally, we show that the passphrase is stored in the volatile memory of the 
device from which it can be extracted and used in the decryption process 
(Sec.~\ref{sec:extraction}).
\subsubsection{The encryption procedure}
\label{sec:encryption}
Before discussing how the secret key can be decrypted, it is necessary
to illustrate the procedure used by ChatSecure to generate, encode, and store
it.
This procedure has been
reconstructed by analyzing the source code of ChatSecure (in particular, file
\textit{WelcomeActivity.java}) and of CacheWord (in particular,
files \emph{PassphraseSecrets.java} and \emph{PassphraseSecrectsImpl.java}),
and is reported in Algorithm~\ref{alg:encryption} below using pseudo-code,
which is executed only when ChatSecure is used for the first time.

\begin{algorithm}[h]
	\caption{ChatSecure secret key generation, encryption, and storage algorithm.}
	\label{alg:encryption}
	\begin{small}
		\begin{algorithmic}[1]	
			\STATE $\mathit{secretKey} = \mathrm{AES.generateSecretKey}(256)$ \label{generate}
			\STATE $\mathit{passPhrase} = \mathrm{readPassPhraseFromUser}()$ \label{passphrase}
			\STATE $\mathit{salt} = \mathrm{generateRandomSalt}()$ \label{salt}
			\STATE $\mathit{IV} = \mathrm{generateRandomInitializationVector}()$ \label{iv}
			\STATE $\mathit{IC} = \mathrm{computeIterCount}()$ \label{icount}
			\STATE $\mathit{passPhraseKey} = \mathrm{pbkdf2}(\mathit{passPhrase},\mathit{salt},\mathit{IV},\mathit{IC})$ \label{derived}
			\STATE $\mathit{encryptedSecretKey} = \mathrm{AES.encrypt}(\mathit{secretKey},\textit{passPhraseKey},\textit{IV})$\label{encrypt}
			\STATE $\mathit{serializedSecret} = \mathrm{concatenate}(\mathit{IC},\mathit{salt},\mathit{IV},\mathrm{base64Encode}(\mathit{encryptedSecretKey}))$ \label{serialize}
			\STATE $\mathrm{save}(\mathit{serializedSecret},\mathit{info.guardianproject.cacheword.prefs.xml})$ \label{save}
		\end{algorithmic}
	\end{small}
\end{algorithm}

As shown in Algorithm~\ref{alg:encryption}, a 256-bit key is generated first (line~\ref{generate}),
and then the user is asked to provide a \textit{passPhrase} (line~\ref{passphrase}).

Starting from this passphrase, a 256-bit \emph{derivate key}) named \textit{passPhraseKey} in
Algorithm~\ref{alg:encryption}) is computed (line~\ref{derived})
by means of the \emph{Password-Based Key Derivation Function 2 (PBKDF2)}~\citep{pbkdf2} algorithm.
This latter algorithm requires four distinct parameters, namely the passphrase and
three additional values,
namely a randomly-chosen 128-bit \textit{salt} (line~\ref{salt}),  a randomly-chosen 96-bit 
\textit{initialization vector (IV)}, and a 32-bit integer \textit{iteration counter (IC)} computed
as function of the speed of the processor of the device.

Then, the derivate key \textit{passPhraseKey} is used to encrypt the
secret key used for database encryption (line~\ref{encrypt}) with AES-256, 
and the result is stored in the \textit{encryptedSecretKey} variable.
Finally, the values of \textit{IC}, \textit{salt}, and \textit{IV}
are concatenated with the Base64 encoding of \textit{encryptedSecretKey}, and are saved 
(as a sequence of bytes) into the  \linebreak\textit{info.guardianproject.cacheword.prefs.xml} file.

\begin{figure}[hbtp]
	\centering
	\includegraphics[scale=0.8]{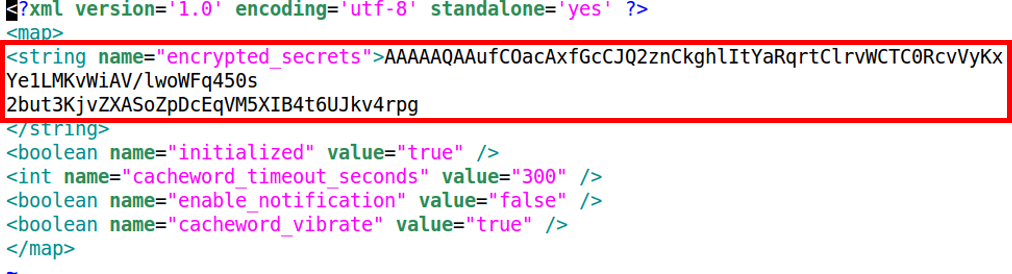}
	\caption{The \emph{serializedSecret} generated and saved by Algorithm~\ref{alg:encryption}.}
	\label{encryption-key}
\end{figure}

An example of the resulting \textit{serializedSecret}
is shown in Fig.~\ref{encryption-key}, where it is highlighted using a square box
drawn around it.

\subsubsection{Decrypting the SQLCipher encryption key}
\label{sec:decrypt-key}
To decrypt the ChatSecure databases, the secret key used with SQLCipher must be known.
Given that this key is unknown to the user, it must be decrypted from
the \textit{serializedSecret} stored in
the \linebreak\textit{info.guardianproject.cacheword.prefs.xml} file.

Assuming that the user passphrase is known, this decryption can be carried
out by means of Algorithm~\ref{alg:decryption}, that we devised starting from
Algorithm~\ref{alg:encryption}.
\begin{algorithm}[h]
	\caption{ChatSecure secret key decryption algorithm.}
	\label{alg:decryption}
	\begin{small}
		\begin{algorithmic}[1]
			\STATE $\mathit{passPhrase} = \mathrm{getPassPhrase}()$ \label{d-passphrase}
			\STATE $\mathit{serializedSecret} = \mathrm{readFromFile}(\mathit{info.guardianproject.cacheword.prefs.xml})$ \label{d-serialized}
			\STATE $\mathit{IC} = \mathrm{extractFromSequence}(\mathit{serializedSecret},0)$ \label{d-icount}
			\STATE $\mathit{salt} = \mathrm{extractFromSequence}(\mathit{serializedSecret},32)$ \label{d-salt}
			\STATE $\mathit{IV} = \mathrm{extractFromSequence}(\mathit{serializedSecret},160)$ \label{d-iv}
			\STATE $\mathit{encryptedSecretKey} = \mathrm{extractFromSequence}(\mathit{serializedSecret},256)$ \label{d-encrypt}
			\STATE $\mathit{passPhraseKey} = \mathrm{pbkdf2}(\mathit{passPhrase},\mathit{salt},\mathit{IV},\mathit{iterCount})$ \label{d-derived}
			\STATE $\mathit{decodedSecretKey} = \mathrm{base64Decode}(\mathit{encryptedSecretKey})$ \label{d-decode}
			\STATE $\mathit{decryptedSecretKey} = \mathrm{AES.decrypt}(\mathit{decodedSecretKey},\mathit{passPhraseKey},\mathit{IV})$ \label{decrypt}
		\end{algorithmic}
	\end{small}
\end{algorithm}

To decrypt the secret AES key from the \textit{serializedSecret}, first the
user-generated \textit{passphrase} is obtained in some way (either by the user or, as
discussed in Sec.~\ref{sec:extraction}, by extracting it from the volatile memory of the device).
Then, the \textit{serializedSecret} is read from the \linebreak\textit{info.guardianproject.cacheword.prefs.xml}
file (line~\ref{d-serialized}), and is subsequently decomposed into its constituent elements, 
namely \textit{IC}, \textit{salt}, \textit{IV}, and 
\textit{encryptedSecretKey} (lines~\ref{d-icount}--\ref{d-encrypt}).
The second parameter of function \textit{extractFromSequence} indicates the offset (expressed in bits)
from the beginning of the \textit{serializedSecret} sequence where each element is stored
(and it is computed by considering the size of each component).

To decrypt \textit{encryptSecretKey}, the derived key \textit{passPhraseKey}
used to encrypt it (see Algorithm~\ref{alg:encryption}, line~\ref{encrypt})
is computed first by means of the PBKDF2 function (line~\ref{d-derived})
using the same values of \textit{salt}, \textit{IV}, and \textit{IC} used
to generate it in Algorithm~\ref{alg:encryption} (line~\ref{d-derived}),
as well as the \textit{passPhrase}.

Then, to obtain the SQLCipher encryption key, we first Base64-decode
the value stored in \textit{encryptedSecretKey}
(recall that in Algorithm~\ref{alg:encryption} this key is
Base64-encoded before being stored, see line~\ref{serialize}).
The result of this operation is stored in variable \textit{decodedSecretKey}
(line~\ref{d-decode}), which is finally decrypted to yield the SQLCipher key
\textit{decryptedSecretKey} (line~\ref{decrypt}).

As an example, the decryption of  the \textit{serializedSecret} shown in Fig.~\ref{encryption-key} yields
the SQLCipher key `\textit{62 9B 8D BF 3F 26 13 1B 2F B6 96 19 FD 4C F9 92 A1 D2 D0 12 96 B5 73 BA 34 59 FA FF 8A 12 CD 89}' (expressed
as a sequence of bytes in hexadecimal encoding).

We have implemented the above decryption algorithm as an Android app that
exploits parts of the CacheWord source code 
(in particular, file \emph{PassphraseSecrectsImpl.java}),
which is freely available upon request.
The choice of implementing it for Android and not for another
platform stems from the fact that the CacheWord source code
does not correctly compile outside the Android development
environment.

\subsubsection{Decrypting ChatSecure databases}
\label{sec:decrypt-db}
Once the encryption key used with SQLCipher has been obtained
by means of Algorithm~\ref{alg:decryption}, the ChatSecure databases
can be decrypted using any SQLite v.3 client that supports SQLCipher.

\begin{figure}[hbtp]
	\centering
	\includegraphics[scale=0.70]{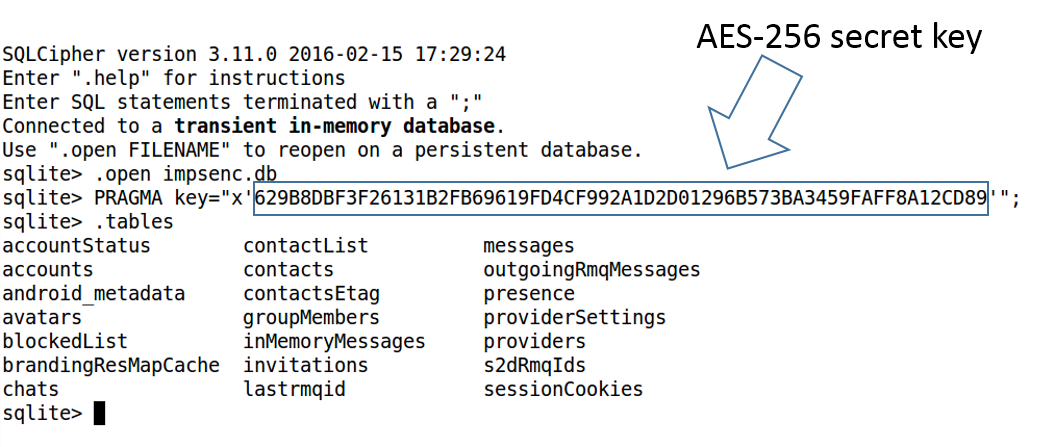}
	\caption{Decrypting ChatSecure \textit{impsenc.db} with SQLCipher.}
	\label{impsenc-decrypt}
\end{figure}

In Fig.~\ref{impsenc-decrypt} we show how the main database
\textit{impsenc.db} can be decrypted on a Linux system by means of the SQLCipher command line tool
(freely available from~\citep{sqlcipher}).

After launching the SQLCipher client, the encrypted database
is opened first by means of the \textit{.open impsenc.db} command.
Then, it is decrypted by means of the  \textit{PRAGMA key = ``x`KEY\_BYTES{'}'';}
command, where \linebreak\textit{KEY\_BYTES} denotes the hexadecimal encoding of the
sequence of bytes corresponding to encryption key.
The last \textit{.tables} command shown in Fig.~\ref{impsenc-decrypt}
serves only to verify that decryption has been correctly performed, as 
in the case of a wrong key the PRAGMA directive fails silently.

\begin{figure}[hbtp]
	\centering
	\includegraphics[scale=0.70]{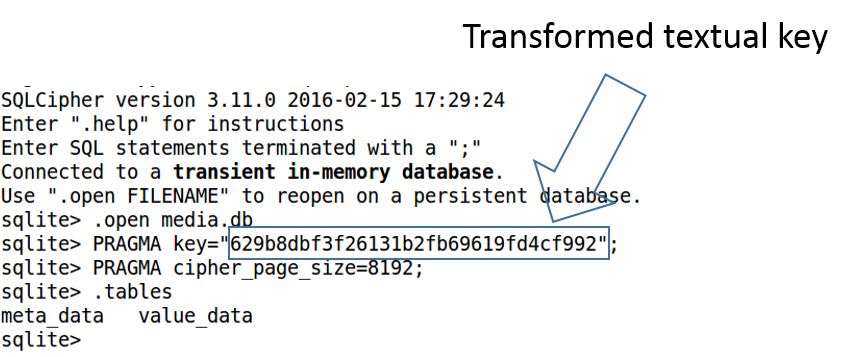}
	\caption{Decrypting ChatSecure \textit{media.db} with SQLCipher}
	\label{media-decrypt}
\end{figure}

The decryption procedure for the \textit{media.db} database
is slightly different, as shown in Fig.~\ref{media-decrypt}.
In particular, a textual key is used in place of the 256-bit SQLCipher key used
for the \textit{impsenc.db} database (where the key was passed to the \textit{PRAGMA key} command
as a hexadecimal sequence).
This textual key is obtained by first converting the 256-bit SQLCipher key
into a lower-case textual string (by translating each hexadecimal digit into
the corresponding ASCII character), and then by truncating it to the leftmost 32 characters.
After this key has been computed, the \textit{media.db} is decrypted by
means of the \textit{PRAGMA key=``TXT\_KEY''} command (where \textit{TXT\_KEY} denotes it)
followed by the \textit{PRAGMA cipher\_page\_size = 8192;} command~\footnote{The
value of $8192$ for the page size for the encrypted database is a design choice of the libsqlfs library (see function \textit{sqlfs\_t\_init} in the
\textit{sqlfs.c} file of the libsqlfs source tree).}
(that, instead, was not required for the \textit{impsenc.db} database),
as shown in Fig.~\ref{media-decrypt}.

\subsubsection{Extracting the passphrase from volatile memory}
\label{sec:extraction}
As discussed before, Algorithm~\ref{alg:decryption} needs the user-defined passphrase 
to decrypt the SQLCipher encryption key, that must be gathered in order to proceed
with database decryption and analysis.

If the ChatSecure user is unwilling to reveal the passphrase,
this problem becomes hard to solve, since this passphrase is never
stored on the persistent memory of the device, so it cannot be retrieved from there.
However, as discussed below, the passphrase persists in the volatile
memory of the device after it has been inserted by the user when
ChatSecure is started. Therefore, if the device is switched on
and ChatSecure is running, the passphrase can be located in the
volatile memory, and can be extracted from there.

In this section, we first discuss how we found out that the
passphrase persists in volatile memory, and then we show how
it can be identified and extracted from a dump of its contents.

To verify whether the passphrase persists in the volatile memory of the
ChatSecure
device, we performed experiments in which we started ChatSecure, entered the passphrase,
put the application in the background,
waited a given amount of time during which the app was not used;
then, we extracted 
the contents of volatile memory of the ChatSecure process, and searched it for the
passphrase that was entered.
Experiments were organized in rounds, where each round included
experiments in which we progressively increased the amount of time we
waited before performing acquisition, up to a maximum of two hours.
We ran different sets of rounds, each one corresponding to a different
passphrase.
Memory extraction and analysis was carried out by using the methodology
described in \citep{lime-howto}, using \textsf{LiME} for extraction and
\textsf{Volatility} for analysis.

\begin{figure}[hbtp]
	\centering
	\includegraphics[scale=0.8]{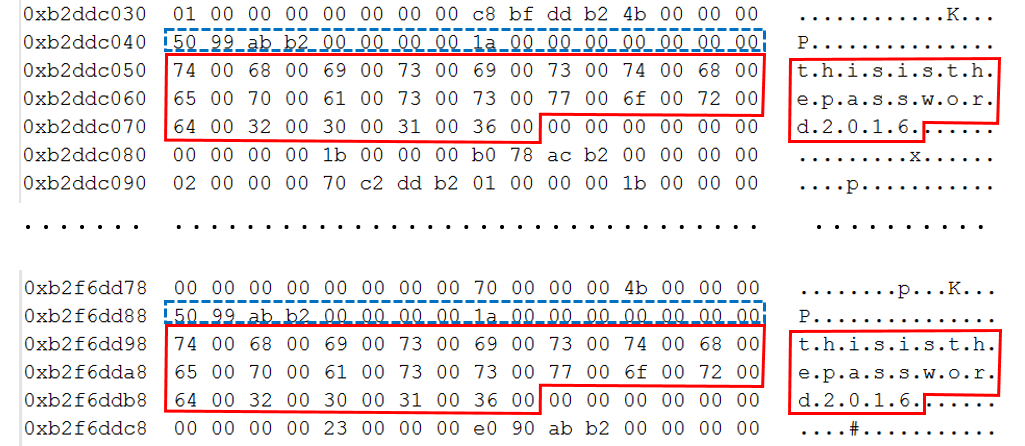}
	\caption{Passphrase in the volatile memory of the device.}
	\label{pass-in-memory}
\end{figure}

The results of our experiments can be  summarized as follows:
\begin{enumerate}
	\item the passphrase was always found in the volatile memory of the ChatSecure process,
	thus proving that it persists there for the entire execution
	of ChatSecure;
	\item the passphrase is stored as a null-terminated Unicode UTF16-LE 
	string (an example is shown in Fig.~\ref{pass-in-memory} for the passphrase
	\textit{thisisthepassword2016}, which is highlighted by continuous-line box surrounding it);
	\item the sequence of bytes encoding the passphrase is preceded by the
	16-bytes signature
	\textit{50 99 ab b2 00 00 00 00 1a 00 00 00 00 00 00 00} (highlighted in Fig.~\ref{pass-in-memory}
	by a dotted-line box surrounding it);
	\item the passphrase and its signature appear twice in the memory space of the
	ChatSecure process, as exemplified in Fig.~\ref{pass-in-memory}.
\end{enumerate}

Of course in real cases the passphrase is not known, so it cannot be
found in memory by simply searching for it, as we instead did in our
experiments.
We need therefore to define a method allowing us to identify an unknown
passphrase stored in volatile memory.

A natural choice would be that of using the 16-bytes signature mentioned above
as a landmark indicating the position of the passphrase.
Unfortunately, the results of our experiments indicate that this method
yields a large number of false positives, since the above signature is present also for other
UTF16-LE null-terminated strings.

However, we can leverage the fact that the passphrase appears \emph{twice}
in the volatile memory of the ChatSecure process, each time preceded by
the above 16-bytes signature, to prune all the candidate strings that do not
occur twice in the above memory space (they are clearly false positives).
Although we cannot exclude that this procedure will filter out all the false positives,
it is certainly able to greatly reduce their number.

Finally, after all the candidate passphrase have been extracted from the volatile
memory region belonging to the ChatSecure process, we can find the correct one
by first running Algorithm~\ref{alg:decryption} for each one of them, and then by attempting
to decrypt the ChatSecure databases with the secret key it returns.
It is worth noticing that the database decryption procedure can be implemented
by using the SQLCipher API, thus making the above method fully automatable.
\subsection{Dealing with deletions}
\label{sec:deletion}
The last issue we consider is concerned with the recovery of messages and
files deleted by the ChatSecure user.
These deletions are performed by deleting the corresponding records 
from the \textit{impsenc.db} and \textit{media.db} databases.

It is well-known that in SQLite databases deleted records are kept in the
so-called  \emph{unallocated cells}, i.e.\ slack space stored in the file 
corresponding to the database, from which they can be recovered~\citep{sqlite-deletion}.

Unfortunately this is not the case for ChatSecure databases, since their
records are deleted securely, i.e.\ they are overwritten upon deletion.
As a matter of fact, in SQLCipher (that, as already discussed, is used 
both by ChatSecure and IOCipher to encrypt the \textit{impsenc.db} and the 
\textit{media.db} databases, respectively), secure deletion is enabled
by default, as reported in its official documentation~\citep{sqlcipher-deletion},
that states:
\begin{quote}
	``\emph{(\dots) as of version 2.0.5, SQLCipher now enables  SQLite's PRAGMA secure\_delete=ON option. This 
		causes the freed  pages to be zeroed out on delete to hinder recovery. As before, they  remain encrypted.
		Note that this doesn't imply that the pages are  removed from the database file, just that their content is 
		wiped when  they are marked free}.'' 
\end{quote}

To verify whether the above holds true in reality, we performed a set of experiments
in which we deleted various messages and files from the above databases, and then
we attempted to recover the corresponding records by means of specialized tools~\citep{ufed_physical,oxygen-sqlite}.
Our analysis did not yield any result, thus indicating that secure deletion is
actually working in the current version of ChatSecure.

We have therefore to conclude that the recovery of deleted messages and files is not possible.
\section{Conclusions}
\label{conclusions}
In this paper we have
discussed the forensic analysis of ChatSecure, a
secure IM application that adopts strong encryption for transmitted and 
locally-stored data to ensure the privacy of its users.

In particular, we have shown that ChatSecure stores local copies of both exchanged
messages and files into two distinct databases, that are strongly encrypted
by means of the SQLCipher library.
Although the encryption mechanisms used by ChatSecure is rather
complex, we have devised an algorithm able to decrypt these
databases starting from the secret passphrase chosen
by the user as the initial step of the encryption process.

We have also shown how this passphrase can be identified and extracted
from the volatile memory of the device, where it persists -- after having
been entered by the user -- for the entire execution of ChatSecure,
thus allowing one to carry out decryption even if the passphrase is
not revealed by the user.

Moreover, we have also shown how to analyze and correlate the
data stored in the databases used by ChatSecure to identify the
IM accounts used by the user and his/her buddies to communicate,
as well as to reconstruct the chronology and contents 
of the messages and files that have been  exchanged among them.

Finally, we have shown that the data stored in the databases cannot be recovered
after having been deleted, as a consequence of the secure deletion technique 
adopted by SQLCipher.

The study reported in this paper has been performed by means of a methodology
that is based on the use of emulated devices and therefore  provides a very high
degree of reproducibility of the results. The accuracy of the method has been
assessed by validating the results it yields against those obtained from
real smartphones.
We believe that this methodology represents also a significant contribution
of this paper.
\bibliographystyle{model4-names}

\end{document}